\documentclass{article}
\usepackage{ijcai23}

\usepackage{times}
\usepackage{soul}
\usepackage{url}
\usepackage[hidelinks]{hyperref}
\usepackage[utf8]{inputenc}
\usepackage[small]{caption}
\usepackage{graphicx}
\usepackage{amsmath}
\usepackage{amsthm}
\usepackage{booktabs}
\usepackage{algorithm}
\usepackage{algorithmic}
\usepackage[switch]{lineno}

\usepackage{color}
\usepackage[english]{babel}
\usepackage{amsthm}
\usepackage{mathtools}
\usepackage{mathrsfs}
\usepackage{amssymb}
\usepackage{bbm}
\usepackage{multirow}
\usepackage{array}
\usepackage{booktabs}
\usepackage{bbding}
\usepackage{dsfont}
\usepackage{makecell}

\title{GreenFlow: A Computation Allocation Framework for Building Environmentally Sound Recommendation System}

\author{
Xingyu Lu$^{1}$\footnote{Both authors contributed equally to this research.} \and
Zhining Liu$^{1*}$\and
Yanchu Guan$^1$\and
Hongxuan Zhang$^2$\footnote{Work done at Ant Group.} \and 
Chenyi Zhuang$^1$\footnote{Corresponding author: chenyi.zcy@antgroup.com.}\and
Wenqi Ma$^1$\and
Yize Tan$^1$\and
Jinjie Gu$^1$\And 
Guannan Zhang$^1$
\affiliations
$^1$Ant Group, $^2$Nanjing University\\
}

\begin{document}

\maketitle

\begin{abstract}
Given the enormous number of users and items, industrial cascade recommendation systems (RS) are continuously expanded in size and complexity to deliver relevant items, such as news, services, and commodities, to the appropriate users. In a real-world scenario with hundreds of thousands requests per second, significant computation is required to infer personalized results for each request, resulting in a massive energy consumption and carbon emission that raises concern. 

 This paper proposes GreenFlow, a practical computation allocation framework for RS, that considers both accuracy and carbon emission during inference. For each stage (e.g., recall, pre-ranking, ranking, etc.) of a cascade RS, when a user triggers a request, we define two actions that determine the computation: (1) the trained instances of models with different computational complexity; and (2) the number of items to be inferred in the stage. We refer to the combinations of actions in all stages as \textit{action chains}. A reward score is estimated for each action chain, followed by dynamic primal-dual optimization considering both the reward and computation budget. Extensive experiments verify the effectiveness of the framework, reducing computation consumption by 41\% in an industrial mobile application while maintaining commercial revenue. Moreover, the proposed framework saves approximately 5000kWh of electricity and reduces 3 tons of carbon emissions per day.
\end{abstract}

\section{Introduction}
With the rapid development of AI, especially after the 2012 breakthrough, i.e., the deep learning revolution, computation has grown substantially. The computations required for deep learning research have been doubling every few months, resulting in an estimated 300,000x increase from 2012 to 2018~\cite{openai_ai_and_compute_2018}. This significant increase in computation has led to a huge energy consumption and carbon emissions. Facing the explosive growth of the demand for computation, Green AI~\cite{Schwartz2019GreenA,Xu2021ASO,Yigitcanlar2021GreenAI}, which aims to maximize energy efficiency and minimize environmental impact, has attracted a lot of research interest.

\begin{figure}[h!]
    \centering
\includegraphics[width=0.32\textwidth]{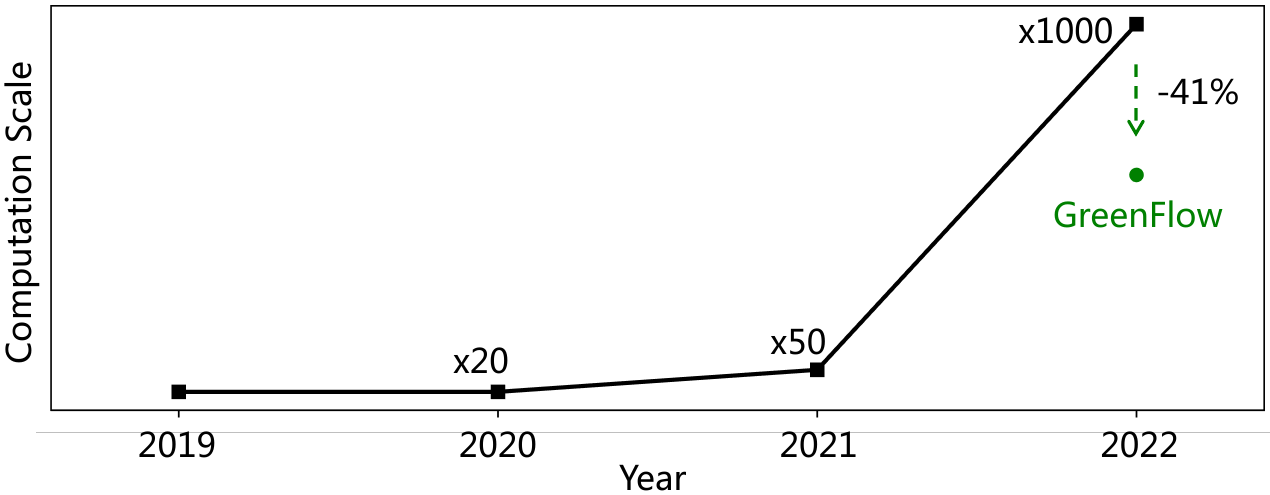}
    \caption{The growth trend of computation consumed by RSs on an industrial mobile application. With GreenFlow, the growth trend has been substantially slowed down.}
    \label{fig:growth_trend}
\end{figure}

RS serves as one of the most widely used AI-based application in various domains, such as content recommenders for multimedia services~\cite{Covington2016DeepNN}, product recommenders for e-commerce~\cite{Zhou2017DeepIN}, or tweet recommenders for social media platforms~\cite{Chen2012CollaborativePT}. Due to large-scale and wide-range of RS, it contributes to a large percentage of AI computation. As shown in Figure~\ref{fig:growth_trend}, we show the growth trend of computation in recent years for RSs of an industrial mobile application. The rapidly growing curve poses a great challenge to the computation overhead. 
However, we found that the difficulty of identifying users' preferences is significantly different, which suggests that the demand for computation of each user is different and an equal allocation strategy would easily lead to an inefficient use of computation.
In addition, industrial RSs are often equipped with multiple cascade AI models of varying levels of computation to satisfy the need for low latency (typically within a few hundred milliseconds) and limited computation budget. However, the computation budget allocated to each stage is usually fixed, which also results in inefficiencies.

To maximize the utility of computation consumed by the RS, two critical problems arise in computation allocation: (1) how to estimate the reward curve for each user given different levels of computation; (2) how to allocate the limited computation to arriving requests effectively based on personalized demand. To address these challenges, we propose a computation allocation framework for the industrial cascade RS as shown in Figure~\ref{fig:system_overview}. Specifically, we first define the two actions that determine the computation as (1) the trained instances of models with different computational complexity; and (2) the number of items to be inferred in each stage. Moreover, we design action chains as the allocation unit, which denote the combinations of model instances and the number of items of all the stages. Next, we estimate a reward score and computation cost for each action chain, followed by a dynamic primal-dual optimization considering both the reward and computation budget. This approach results in an optimal action chain for computation allocation.

\noindent
\textbf{Scientific Contribution.}
In summary, our contributions mainly include the following three aspects:
\begin{itemize}
\item \textbf{Problem.} We formally define the problem of computation allocation with the goal to maximize its revenue given a limited computation budget in an industrial RS.
\item \textbf{Method.} We propose a generic computation allocation framework, which maximizes the revenue of computation through (1) building an personalized reward estimation model of action chains, and (2) conducting dynamic primal-dual optimization for online allocation.
\item \textbf{Evaluation.} We perform extensive experiments showing that GreenFlow consistently achieves improvements. By deploying GreenFlow on several industrial RSs, we are able to save approximately 5000kWh of electricity and reduce carbon emissions by 3 tons per day.
In addition, we present an evaluation methodology named PFEC, which thoroughly evaluates the effectiveness of the computation allocation method from four aspects: \textbf{P}erformance, \textbf{F}LOPs, \textbf{E}nergy, and \textbf{C}arbon emission.
\end{itemize}

\noindent
\textbf{Impact to the SDGs.} We believe that such a computation allocation framework can greatly improve the resource-use efficiency (SDG 9), which can make the RS environmentally sound and also contributes a lot to the combat climate change (SDG 13) because it can greatly reduce carbon emission without harming the commercial revenue.

\section{Related Work}
\subsection{Green AI}
Green AI is an emerging research field that aims to develop strategies to reduce energy consumption and mitigate the climate impact of AI systems to maximize energy efficiency. We summarize its research directions as follows:
\begin{itemize}
    \item Data center: This direction mainly focuses on developing more energy-efficient algorithms and hardware, optimizing data center cooling and power usage~\cite{Khosravi2017DynamicVP}, and using renewable energy sources to reduce the power usage effectiveness (PUE).
    \item Computing cluster and engine: It focuses on improving the conversion efficiency of energy to computation via optimizing the scheduler~\cite{Borg}, parallel computing~\cite{Shukur2020ASO}, and other techniques.
    \item AI/ML application: Through optimizing the all stages of deep learning~\cite{Xu2021ASO} towards lower computation costs, the conversion efficiency of computation to revenue is improved. Among existing directions, architecture design~\cite{Kitaev2020ReformerTE,Howard2017MobileNetsEC}, training~\cite{Liu2021SamplingMF,Huang2018GPipeET} and inference~\cite{Molchanov2016PruningCN,Han2016EIEEI} are major research directions.
\end{itemize}
In this paper, we mainly focus on RS from the perspective of AI/ML application, and achieve the goal of Green AI via allocating the computation in an efficient way.

\subsection{Computation Allocation in RS}
Multi-stage cascade ranking systems is a common solution in the industry~\cite{Wang2011ACR,Qin2022RankFlowJO} to achieve ranking of a great amount of items with low latency. Its main idea is to deploy simple rankers in the early stages for pre-ranking on the scale of the entire candidate set, and use complex rankers for a much smaller set of items for better accuracy, so that we can achieve a well trade-off between efficiency and effectiveness. 
To our knowledge, \cite{jiang2020dcaf} is the first paper to propose the allocation of computation in RS at the granularity of online requests, where the size of candidate items is adjusted for each user. However, this approach only focuses on one specific stage of RS and may not perform well on a cascade architecture. \cite{Yang2021ComputationRA} extends the method to cascade RS, assuming that the revenue of each stage is independent of the others and can be multiplied to obtain the total revenue of the cascade RS.

However, these studies do not consider the joint effects across stages and different models, which results in a failure to model the personalized reward model at a fine-grained level and allocate the computation effectively.

\begin{figure*}[h!]
    \centering
    \includegraphics[width=0.96\textwidth]{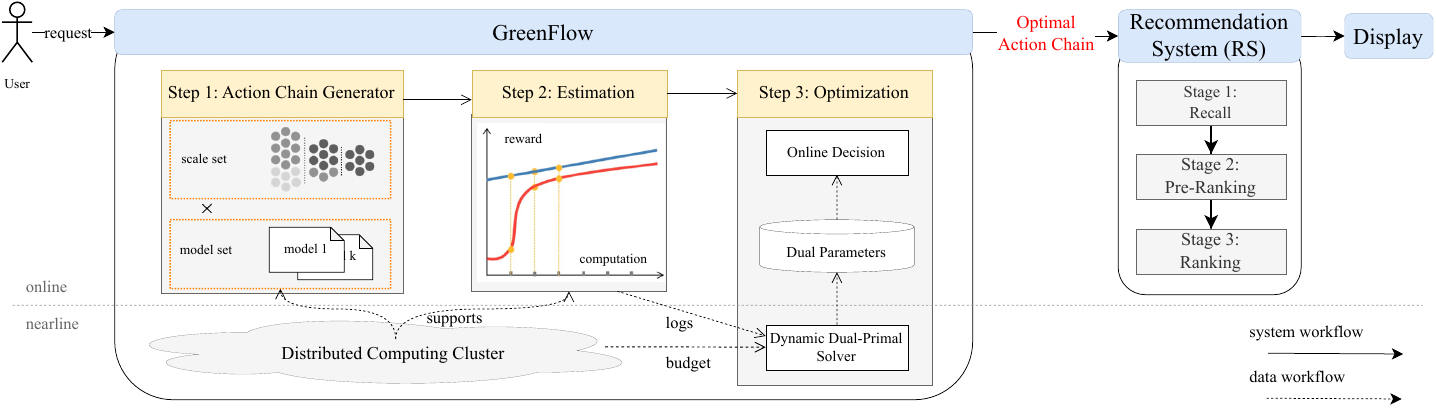}
    \caption{System overview of the proposed GreenFlow.}
    \label{fig:system_overview}
\end{figure*}

\section{Framework}
\subsection{Overview}
In a multi-stage cascade RS, we introduce the GreenFlow framework to maximize the efficiency of the whole RS under the limited computation budget. Figure \ref{fig:system_overview} illustrates the whole process of a RS after introducing the GreenFlow. In order to take computation budget into consideration, we introduce the new concept called \textit{action chain}. Since the computation cost of a certain ranking stage is determined by the computational complexity of ranking models and the number of items to be inferred, so that we adopt an action chain to assemble them for all stages. Then, GreenFlow aims to allocate an optimal action chain for each incoming request prior to the RS. In details, the framework is consist of three steps. 

In the first step, the \textit{action chain generator} constructs candidate action chains through cartesian product between models and item scales. 

Then, in the second step, we develop a computation measure module and a reward model to estimate the computation costs and rewards of each candidate action chain, respectively. For a comprehensive computation cost estimation, several metrics, such as FLOPs, carbon emissions, and CPU usage, are calculated using static tools (e.g.,  experiment-impact-tracker
\footnote{https://github.com/Breakend/experiment-impact-tracker}, carbontracker\footnote{https://github.com/lfwa/carbontracker}, etc.). For the reward estimation task, in the context of a specific request, our model uses the actual revenues (e.g., click rate, conversion rate, etc.) as labels for training. In general, an action chain with a higher computation cost tends to generate better revenue. However, it should be noted that the uplift of the reward curve varies with (1) different requests and (2) the corresponding stages as the computation cost increases.

In the third step, we propose a hybrid online-nearline architecture to obtain the optimal action chain for allocation under the global computation budget. In the nearline module, a streaming job collects the candidate action chains' reward scores and computation costs of each request. Then, a dynamic primal-dual solver is proposed to do a matching task using the collected samples and the current budget periodically. Finally, the solved dual parameters are stored into an online storage. The entire procedure of updating the dual parameters takes minutes or seconds, depending on the response time requirements of the RS. In the online module, the framework returns an optimal action chain for each arriving request to the RS.

However, it should be noted that introducing GreenFlow itself into RS will result in additional computation. We will further discuss and quantify this trade-off in the experiment section.

\subsection{Evaluation Methodology}
To quantitatively measure the effectiveness of the computation allocation framework in increasing computation utilization and combating climate change, we propose an evaluation methodology named \textbf{PFEC}, which thoroughly evaluates the effectiveness of the framework from the following four aspects, i.e., \textbf{P}erformance, \textbf{F}LOPs, \textbf{E}nergy, and \textbf{C}arbon emission.

Performance is typically used to measure the revenue generated by RS, with common metrics including clicks, conversions, and other similar measures. FLOPs, which measures how many operations are required to run a single instance of a given neural network, can quantify how much the computation RS consumes.

As for energy and carbon emission, following the method proposed in \cite{Lacoste2019QuantifyingTC}, we first calculate the energy consumption (EC) mainly taking the RAM, CPU and GPU into consideration:
\begin{equation}
    \text{EC} = \text{PUE} \cdot (p_{\text{ram}}e_\text{ram} + p_\text{cpu}e_\text{cpu} + p_\text{gpu}e_\text{gpu}),
\end{equation}
where $\text{PUE}$ is a constant determined by the energy efficiency of a data center (the worldwide average is 1.67\footnote{https://journal.uptimeinstitute.com/is-pue-actually-going-up/}), $p_{(\cdot)}$ and $e_{(\cdot)}$ denote the rated power of certain device and device usage, respectively. With the $\text{EC}$, we can further calculate the carbon emission (CE) via:
\begin{equation}
    \text{CE} = \text{EC} \cdot \text{CI},
\end{equation}
where $\text{CI}$ is the carbon intensity that varies with different countries, and it is set to 615$gCO_2e\/kWh^{-1}$ in this paper \footnote{https://app.electricitymaps.com/zone}. 

With PFEC, we can quantitatively compare different allocation frameworks on resource utilization and impact on the environment.

\section{Methodology}
\subsection{Notations and Definitions}
In this section, we will introduce the formal definition of the action chain and the computation allocation problem for the whole RS. We assume that $x$ or $X$ denotes a scalar; $\vec{x}$ denotes a vector; $\mathbf{X}$ denotes a matrix; and $\mathcal{X}$ denotes a set. 

Let $\mathcal{I}_t:=\{1,2,...,I_t\}$ be a set of arriving requests in time $t$, $K$ be the number of stages in the cascade RS and $C$ be the computation budget of the whole RS. In each stage $k$, the ranking model $m_k$ and ranking item scale $n_k$ are selected from the \textit{Model Pool} $\mathcal{M}$ and the \textit{Item Scale} set $\mathcal{N}$ as shown in Figure~\ref{fig:system_overview}, respectively. A tuple of $m_k$ and $n_k$ constitute the key parameters $s_k=\{m_k,n_k\}$ of the $k$-th stage. Then, we define the action chain of a whole cascade ranking RS as $a=(s_1, s_2,...,s_K)$ and let $\mathcal{A}$ be the action chain set constructed by the \textit{action chain generator} where $|\mathcal{A}|=J$. For each action chain $a_j$, let $c_j$ be the estimated computation cost. For each request $i$, let $R_{ij}:=R(a_j, f_i)$ be the estimated reward of $a_j$ with the context feature $f_i$ which is related to the arriving request $i \in \mathcal{I}_t$. Then we define the allocation strategy $\mathbf{x}_i:=(x_{i,1},x_{i,2},...,x_{i,J})\in \{0,1\}^J$ that $x_{ij}$ represents the decision variable of allocating action chain $j$ to request $i$. We construct the computation allocation problem as following:  

\begin{subequations}\label{main-problem}
\begin{align}
    & \max_{\mathbf{x}_{i}} \sum_{i\in \mathcal{I}_t,1\leq j\leq J} R_{ij}x_{ij},\\
    s.t. & \quad \sum_{j\leq J} x_{ij} = 1,  \forall i\in\mathcal{I}_{t}, \\
         & \sum_{ i\in \mathcal{I}_t,1\leq j\leq J} c_j x_{ij}\leq C, \\
   x_{ij}&\in\{0,1\}, \forall i\in\mathcal{I}_{t}, 1\leq j\leq J
\end{align}
\end{subequations}

In the next two subsections, we first introduce how to accurately estimate the reward function $R_{ij}$, and then propose a method for quickly solving the Problem~\eqref{main-problem}.

\subsection{Personalized Reward Model}\label{sec:rewardmodel}

\begin{figure}[h!]
    \centering
    \includegraphics[width=0.38\textwidth]{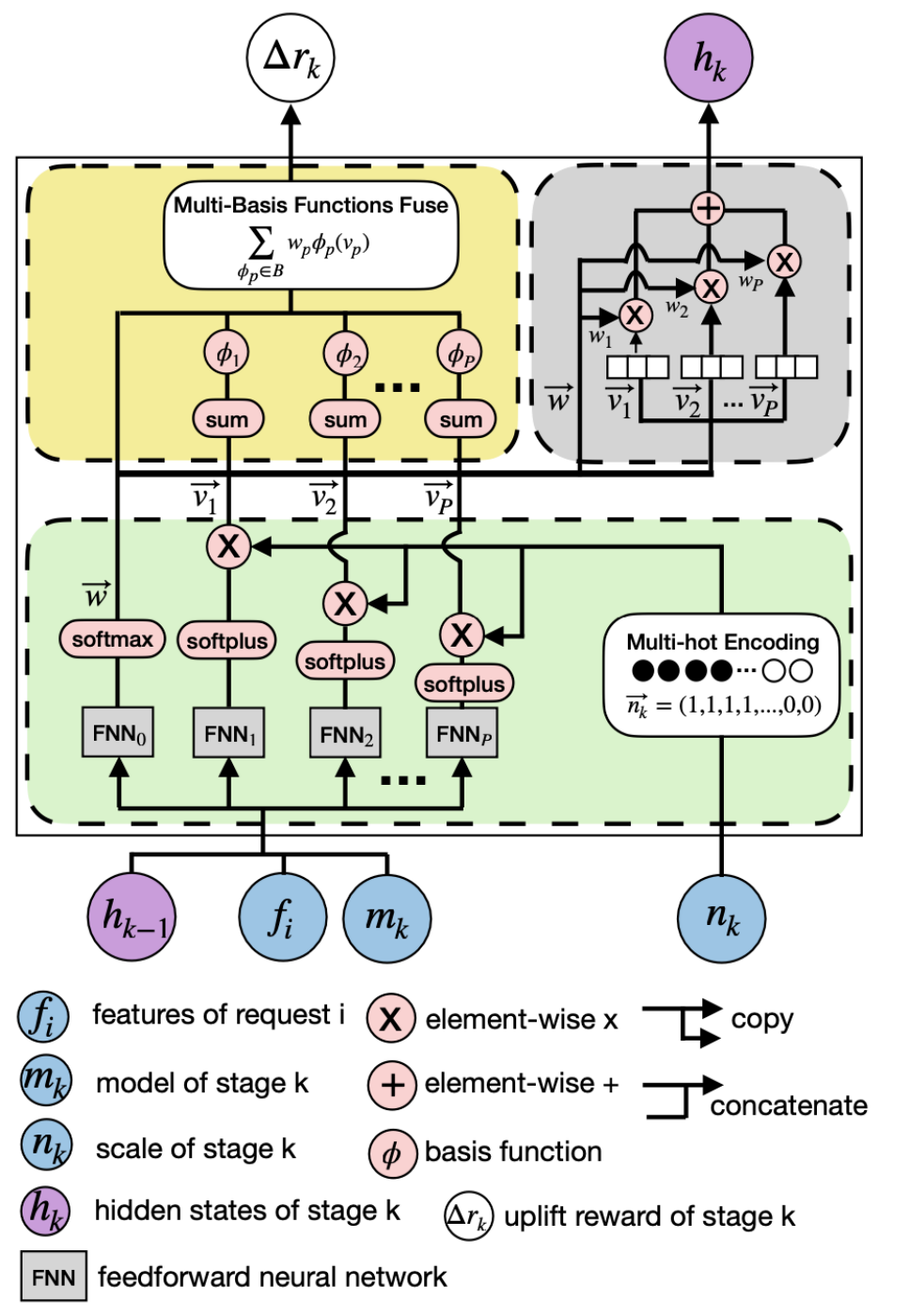}
    \caption{The structure of the recursive function $g_k$.}
    \label{fig:ROI_model}
\end{figure}

In general, increasing computation is likely to result in higher revenue. However, the same action chain may generate different revenues for different requests. Thus, for each request $i$ with context features $f_i$ and action chain $a_j$, it is necessary to design a personalized model to accurately estimate the reward $R_{ij}$. In this regard, we encounter the following three challenges:
\begin{itemize}
\item How to model the dependence of cascade stages in RS?
\item How to deal with the data sparsity problem?
\item How to ensure the monotonicity between computation and reward?
\end{itemize} Accordingly, we introduce the key three mechanisms of the model in the remainder of this section. 

\noindent \textbf{Recursive Multi-Stage Design.} The structure of our proposed model is designed in a recursive style to match the cascade design of RS. As illustrated in Figure \ref{fig:ROI_model}, the chunk of neural network is the detailed implementation of the recursive function $g_k$. Given the output embedding of the previous stage $\vec{h}_{k-1}$, the action $\{m_k, n_k\}$ from $k$-th stage, and the context feature $f_i$, $g_k$ outputs the reward uplift $\Delta r_k $ and $\vec{h}_{k}$ for the next stage. Hence, $R_{ij}$ is calculated by:
\begin{subequations}\label{reward_main}
\begin{align}
    R_{ij} & = \sum_{k=1}^{K}\Delta r_k, \\    
    (\Delta r_k, \vec{h}_k) & = g_k(\vec{h}_{k-1},\vec{f}_i, \vec{m}_k, \vec{n}_k),
\end{align}
\end{subequations} 
where $K$ is the number of stages; $\vec{f}_i, \vec{m}_k, \vec{n}_k$ are the encoded embeddings of $f_i, m_k, n_k$, respectively. Meanwhile, each $g_k$ should deal with data sparsity problem and obey the monotonicity constraint. 

\noindent \textbf{Multi-Basis Functions.} In an industrial RS, collecting feedback for all action chains on the same user can be expensive, which presents a challenge in learning the function $g_k$ using a network with high degrees of freedom. 

So we propose a novel structure of consisting multi-basis functions in the yellow chunks of Figure \ref{fig:ROI_model}. The reward uplift $\Delta r_k$ is given by: 
\begin{equation}
    \Delta r_k = \sum_{\phi_p\in \mathcal{B} }w_p \phi_p(v_p),
\end{equation}
where $\mathcal{B}$ is a set of basis functions and $|\mathcal{B}|=P$; $w_p \in \vec{w}=(w_1, w_2, ..., w_P)$ and $v_{p} \in \vec{v}=(v_{1}, v_{2}, ..., v_{P})$ are two vectors calculated by the feedforward neural networks $\text{FNN}(\cdot )$ in the green chunk of Figure \ref{fig:ROI_model}. The specific form of $\vec{w}$ and $\vec{v}$ can be represented by: 
\begin{equation}\small
\begin{aligned}
    v_p&  = \mathbf{1}_Q^\top (softplus(\text{FNN}_{p}(\vec{h}_{k-1}, \vec{f}_i, \vec{m}_k)) * \vec{n}_k), \forall p\leq P, \\ \label{equation_v_p}
    \vec{w} &  = softmax(\text{FNN}_{0}(\vec{h}_{k-1}, \vec{f}_i, \vec{m}_k)),
\end{aligned}
\end{equation}
where $\mathbf{1}_Q$ is a unit vector, $Q$ is the dimension of the embedding $\vec{n}_k$ and $*$ is the element-wise multiply. In the current implementation, we define the multi-basis functions set as:
\begin{equation}
    \mathcal{B} = \{tanh(x),ln(x),\frac{x}{\sqrt{1+x^2}},sigmoid(x),x\}.
\label{basis-function}
\end{equation} 
This design has the following advantages:

\begin{itemize}
  \item The use of multi-basis functions can better capture the reward curves of users with varying levels of activity.
  \item The design of different basis functions can make the reward model obtain different mathematical properties. For instance, if the second derivative of the basis functions we design is non-positive, then the marginal revenue of the reward curves will be non-increasing. That is, as FLOPs become larger, the slope of the reward curve will become smaller.
\end{itemize}

In the experiment section, we will conduct ablation experiments to demonstrate the advantages of the aforementioned design and the specific effects it achieves.

\noindent 
\textbf{Monotonic Constraint.}
To obey the monotonicity between computation (especially the item scale ${n}_k$) and reward $\Delta r_k$, we use a multi-hot embedding $\vec{n}_k$ for the item scale ${n}_k$ in each stage $k$. In the multi-hot encoding module, the item scale set $\mathcal{N}_k$ is divided into $Q$ groups and each $n_k$ in the same group shares the same embedding. The larger item scale $n_k$ attains more $1$s in $\vec{n}_k$ during the encoding process, and the larger $\vec{v}$ computed by Equation~\eqref{equation_v_p} will be. Since all the basis functions are monotonically increasing and $\vec{w}$ are non-negative, the monotonically increasing relationship between the item scale $n_k$ and reward $\Delta r_k$ is preserved.

\subsection{Dynamic Primal-Dual Optimization}
With known requests $\mathcal{I}_t$, estimated revenues $R_{ij}$ and resource cost $c_j$ in time $t$, one can regard the problem defined in Equation~\eqref{main-problem} as a bipartite matching problem~\cite{aggarwal2011online} following a primal-dual framework. By optimizing this problem, the dual variables can be solved directly. Specifically, the Lagrangian function of Equation~\eqref{main-problem} is given by
\begin{equation}
    L_t(\lambda, \mathbf{X}) = \sum_{i\in \mathcal{I}_t,j\leq J}R_{ij}x_{ij} + \lambda (C - \sum_{i\in \mathcal{I},j\leq J}c_jx_{ij}),
\end{equation}
and the dual form of Equation~\eqref{main-problem} leads to:
\begin{equation}\label{dual_problem}
    \min_{\lambda\ge 0}\max_{X} L_t(\lambda, \mathbf{X}),
\end{equation}
where $\lambda$ is the dual variable. In the literature of matching problem, $\lambda$ is called \textit{dual price} which represents the increased revenues of per resource cost given the budget. With the help of strong duality and K.K.T conditions, we can derive the optimal decision $\vec{x}_{i}$ from solved dual price $\lambda^*$ by: 
\begin{align}\label{decision_rule}
x_{ij}=\left\{\begin{array}{ll} 1, ~&\text{if}~ j = \arg\max_{j} \{R_{ij}-c_j\lambda^*\}\\
0,~&\text{otherwise}.\end{array}\right.
\end{align}

We propose a dual descent algorithm to solve the optimal dual price $\lambda_t$ in time $t$ and apply the $\lambda_t$ to make the online decision $\vec{x}_{i}$ for requests $i\in \mathcal{I}_{t+1}$ periodically. The pseudo-code
of the whole procedure is summarized in Algorithm~\ref{DPDA}. In each time $t$, a gradient descent approach is adopted to solve $\lambda_{t+1}$ by step 6$\sim$8 with global convergence guarantee~\cite{li2020simple}. Moreover, \cite{agrawal2014dynamic} shows that the online decision in time $t+1$ using the previous $\lambda_t$ is near-optimal when the requests adopt a stochastic arriving model and the requests $\mathcal{I}_t$ is enough for solving $\lambda_t$. By adopting a seconds or minutes period of Algorithm~\ref{DPDA} in an industrial RS, our algorithm is effective that the distribution of arriving users is i.i.d. in a short time and the number of requests per second are generally thousands. 
 
\begin{algorithm}
\caption{Dynamic Primal-Dual Algorithm \label{DPDA} }
\begin{algorithmic}[1]
\STATE \textbf{Input:} computation budget $C$, action chain set $\mathcal{A}$, computation cost $c_j$ ($\forall a_j\in \mathcal{A}$), max iterations $L$, step size $\eta$ and initialized dual price $\lambda_0$.\\
\FOR{$t=1,2,\dots$}
\STATE Collect the reward $R_{ij}$ for each request $i\in\mathcal{I}_t$.\\
\STATE Let $\lambda^0_t = \lambda_{t-1}$.
\FOR{$l=0,1,2,...,L$}
\STATE $\text{Update }\vec{x}_{i}, \forall i\in\mathcal{I}_t\text{ given } \lambda^l_t \text{ by Equation~\eqref{decision_rule}}.\label{update_1}$ \\
\STATE $\nabla L = C - \sum_{i\in\mathcal{I}_t,1\leq j\leq J} c_jx_{ij}.\label{update_2}$ \\
\STATE $\lambda^{l+1}_t = \lambda^l_t - \eta \nabla L.\label{update_3}$ \\
\ENDFOR

\STATE Let $\lambda_{t} = \lambda^L_t$, and solve $\vec{x}_{i}$ for $i\in\mathcal{I}_{t+1}$ with $\lambda_{t}$ by Equation~\eqref{decision_rule}.
\ENDFOR
\end{algorithmic}
\end{algorithm}

\section{Experiments}
In this section, we will evaluate the proposed framework through various offline and online experiments to demonstrate its effectiveness. Before introducing the experimental results, we will first describe the experiment settings, including the datasets used, details of training, comparison methods, and evaluation metrics.

\subsection{Experimental Setup}
\noindent
\textbf{Dataset.} For reproducibility, we use one public dataset named Ali-CCP\footnote{https://tianchi.aliyun.com/dataset/408} for offline evaluation. It is a public click and conversion prediction dataset collected from traffic logs of Taobao’s RS~\cite{ma2018entire}, and consists of 85 million samples.
We randomly sample 50\% of users in the dataset for training models in the cascade RS, 25\% for building the validation dataset, 22.5\% for training sample generation of the reward model, and the rest 2.5\% for the final evaluation, which includes 9016 users.

\noindent
\textbf{Implementation of Action Chain.} 
In the experiments, we consider a three-stage (i.e., recall, pre-ranking and ranking) cascade RS with four trained instances of models for allocation, of which detailed statistics are shown in Table~\ref{table:model_instances}.
To simplify the comparison, only one model for each of the first two stages, which are DSSM~\cite{Huang2013LearningDS} and YoutubeDNN~\cite{Covington2016DeepNN} (abbreviated as YDNN for brevity), respectively. DSSM is used for ranking the entire candidate items (its size is $n_1$), and YDNN performs pre-ranking on the top-$n_2$ scored candidate items ranked by DSSM. 
As for the ranking stage, the most two powerful models, DIN~\cite{Zhou2017DeepIN} and DIEN~\cite{Zhou2018DeepIE}, are available for selection to generate top-$n_3$ scored candidate items feed by the previous stage. Finally, we choose top scored items to expose to users. Therefore, an action chain is denoted as $a=(\{\text{YDNN},n_2\}, \{\text{DIN or DIEN},n_3\})$, where $n_2 \in \mathcal{N}_2={[}800, 900, 1000,...,1500{]}$ and $n_3 \in \mathcal{N}_3={[}60, 80, 100,...,200{]}$ in this paper ($\{\text{DSSM}, n_1\}$ is omitted due to its fixed computation cost).

\noindent
\textbf{Evaluation Metrics.} To compare different methods thoroughly, we construct various computation budgets $C$ by simulating different action chains. Given a specific budget constraint $C$, we focus on the following offline metrics:
\begin{equation}
    revenue@e=\sum_{i\in \mathcal{I}_t,j\in \mathcal{P}}\hat{R}_{ij},
\end{equation}
where $\mathcal{P}$ denotes the top-$e$ scored candidate items in the ranking stage generated by an action chain (we set $e=20$ in this experiment), and the reward $\hat{R}_{ij}$ of the user $i$ is set to the number of the clicked items. Since it is difficult to simulate the energy consumption in an offline way, we mainly consider the first two indicators of PFEC, i.e., $revenue@e$ and $C$, in the offline experiments, and a full PFEC evaluation is conducted on the online experiments.

\begin{table}[]
\centering
\setlength{\tabcolsep}{5mm}{
\begin{tabular}{@{}llll@{}}
\toprule
\textbf{Stage} & \textbf{Model} & \textbf{FLOPs} & \textbf{AUC}   \\ \midrule
Recall & DSSM                      & 13K                      & 0.525 \\
Pre-ranking & YDNN                & 123K                      & 0.581 \\
Ranking & DIN                       & 7020K                & 0.639       \\
Ranking & DIEN                      & 7098K                & 0.641      \\ \bottomrule
\end{tabular}
}
\caption{Trained instances of models for allocation of each stage.}
\label{table:model_instances}
\end{table}

\noindent
\textbf{Training Sample Generation of Reward Model.} 
To train the personalized reward model, we simulate different action chains for each user and calculate the reward of each action to build the training dataset.
Specifically, the click-through rate on $\mathcal{P}$ is used to represent the reward (i.e., label) of the action chain $a=(\{\text{YDNN},n_2\}, \{\text{DIN or DIEN},n_3\})$ for supervising the training of the reward model.

\noindent
\textbf{Comparison Methods.} Two methods are compared in the experiments:  
\begin{itemize}
    \item \textbf{EQUAL}: Computation is allocated to each user equally based on a given fixed action chain.
    \item \textbf{CRAS}~\cite{Yang2021ComputationRA}: A method that decomposes the computation allocation into independent sub-problems on each stage of RS.
\end{itemize}

\subsection{Offline Experiments}
In this part, the offline experiments are designed to answer the following questions:
\begin{itemize}
    \item \textbf{Q1}: Does our proposed GreenFlow outperform other approaches on the task of computation allocation?
    \item \textbf{Q2}: Does multi-stage modeling in reward estimation outperform single-stage modeling?
    \item \textbf{Q3}: Is it necessary to provide multi-models in one stage for allocation rather than single-model?
    \item \textbf{Q4}: How does each variant of the proposed reward model contribute to the improvement?
    
\end{itemize}

\subsubsection{Q1: Effectiveness of GreenFlow}
\begin{figure}[t!]
    \centering
    \includegraphics[width=0.42\textwidth]{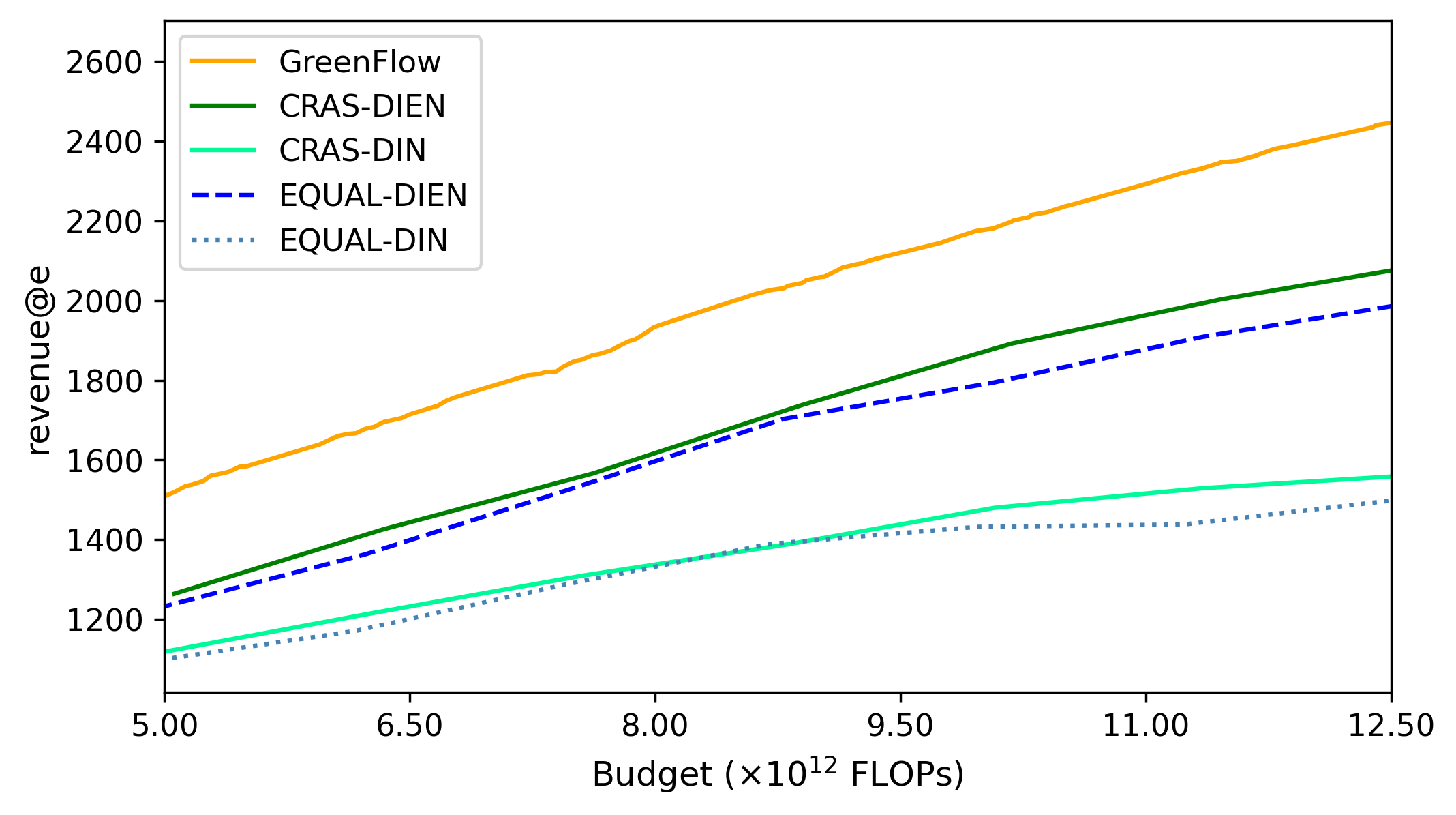}
    \caption{Results of different methods with different budgets.}
    \label{fig:revenue_flops_fig}
\end{figure}

We report the results given different computation budgets in Figure~\ref{fig:revenue_flops_fig}.
Due to the limitation that EQUAL and CRAS only consider single model for allocation in the same stage, we design two variants of EQUAL and CRAS, respectively, i.e. EQUAL-DIN, EQUAL-DIEN, CRAS-DIN and CRAS-DIEN, for comparison as there are two available models in the ranking stage.
From Figure~\ref{fig:revenue_flops_fig}, we can draw the following conclusions: 
\begin{itemize}
    \item With the decreasing budget, the number of clicks is accordingly decreasing, which implies that a well-designed cascaded RS is necessary due to its improvement over the revenue of computation.
    \item Our proposed method outperforms all baseline methods with a large margin, which verifies the superiority of dynamic action chain and multi-stage reward modeling.
\end{itemize}

\subsubsection{Q2: Single-Stage vs. Multi-Stage}
To demonstrate the effectiveness of multi-stage modeling, we compare GreenFlow and CRAS with one action fixed, i.e., our reward model degrades into single-stage. From Table~\ref{table:q2_result}, we observe that CRAS and GreenFlow achieve comparable performance under one action fixed, but GreenFlow outperforms CRAS by a large margin when conducting multi-stage modeling, which verifies the superiority of multi-stage modeling.

\begin{table}[h!]\small
\begin{tabular}{lc|c|cc}
\toprule
\multicolumn{3}{c|}{\textbf{Setup}}                                                                                         & \multicolumn{2}{c}{\textbf{Methods} } \\ 
\midrule

\multicolumn{1}{c|}{Strategy}                      & \multicolumn{1}{c|}{Actions}                                                                        & $\times 10^{12}$FLOPs & \multicolumn{1}{l|}{CRAS} & Ours \\ \midrule
\multicolumn{1}{l|}{\multirow{6}{*}{\makecell[c]{Single-\\Stage}}} & \multirow{3}{*}{\makecell[c]{ \{ $m_3$=DIEN, \\$n_3\in\mathcal{N}_3$\}}}     & 8.68   & 1634                      & \textbf{1665}     \\
\multicolumn{1}{l|}{}                              &                                                                        & 9.96   & 1740                      & \textbf{1784}         \\ 
\multicolumn{1}{l|}{}                              &                                                                        & 11.24  & 1833                      & \textbf{1873}         \\ \cline{2-5}
\multicolumn{1}{l|}{}                              &  \multirow{3}{*}{\makecell[c]{\{$m_2$=YDNN, \\$n_2\in\mathcal{N}_2$\} }}     & 7.33   & 1223                      & \textbf{1248}         \\  
\multicolumn{1}{l|}{}                              &                                                                        & 7.55   & \textbf{1310}                      & 1301         \\  
\multicolumn{1}{l|}{}                              &                                                                        & 7.88   & 1516                      & \textbf{1524}         \\  
\midrule

\multicolumn{1}{l|}{\multirow{6}{*}{\makecell[c]{Multi-\\Stage}}}  & \multicolumn{1}{l|}{\multirow{6}{*}{\makecell[c]{ ( \{$m_2$=YDNN,\\ $n_2\in \mathcal{N}_2$\}, \\ \{ $m_3$=DIEN\\or DIN,\\ $n_3\in\mathcal{N}_3$\})}}} & 8.68       & 1624                          & \textbf{2023}          \\ 
\multicolumn{1}{l|}{}                              & \multicolumn{1}{l|}{}                                        & 9.96       & 1765                          & \textbf{2174}          \\  
\multicolumn{1}{l|}{}                              & \multicolumn{1}{l|}{}                                        & 11.24      & 1845                          & \textbf{2321}          \\  \cline{3-5}
\multicolumn{1}{l|}{}                              & \multicolumn{1}{l|}{}                                        & 7.33       & 1225                          & \textbf{1820}          \\  
\multicolumn{1}{l|}{}                              & \multicolumn{1}{l|}{}                                        & 7.55       & 1309                          & \textbf{1851}          \\  
\multicolumn{1}{l|}{}                              & \multicolumn{1}{l|}{}                                        & 7.88       & 1522                          & \textbf{1903}          \\

\bottomrule
\end{tabular}
\caption{Results with different computation budgets.}
\label{table:q2_result}
\end{table}

\subsubsection{Q3: Single-Model vs. Multi-Model}
As presented in Table~\ref{table:model_instances}, both DIN and DIEN models have similar FLOPs and perform comparably in terms of AUC. This naturally raises the question of whether introducing different models in the same stage is necessary. We argue that this necessity based on the fact that no single model can outperform all others on all users.
To evaluate this, we divided the user set into three groups: those better suited to DIN, those better suited to DIEN, and those equally suited to both. The distribution of these groups is approximately 1:3:6, respectively.
Then, we conducted three tests on GreenFlow using only DIN, only DIEN, and both of them. The results are shown in Table~\ref{table:one_model_study}, and it is evident that using multiple models yields gains ranging from 1$\%$ to 6$\%$ on $revenue@e$ compared to using any single model.

\begin{table}[h!]
\centering
\begin{tabular}{c|ccc}
\toprule
\multicolumn{1}{c|}{\textbf{Budget $C$}}                                                    & \multicolumn{3}{c}{\textbf{Methods}}                                                                                                                                                     \\ \midrule
\multicolumn{1}{p{18mm}|}{$\times 10^{12}$FLOPs}  & \multicolumn{1}{p{18mm}|}{Only DIN}   & \multicolumn{1}{p{18mm}|}{Only DIEN}   & \multicolumn{1}{p{16mm}<{\centering}}{Both} \\ \midrule
6.2                                                 & 1345                                  & \multicolumn{1}{c|}{1572}         & \textbf{1667}                  \\
7.4                                                 & 1467                                  & \multicolumn{1}{c|}{1814}         & \textbf{1834}                  \\
7.5                                                 & 1471                                  & \multicolumn{1}{c|}{1823}         & \textbf{1848}                  \\
8.7                                                 & 1581                                  & \multicolumn{1}{c|}{2007}         & \textbf{2026}                  \\
10.1                                                & 1686                                  & \multicolumn{1}{c|}{2079}         & \textbf{2181}                  \\
\bottomrule
\end{tabular}
\caption{Results of GreenFlow with different models available for selection in the ranking stage.}
\label{table:one_model_study}
\end{table}

\subsubsection{Q4: Variants of Reward Modeling} To verify the effectiveness of the proposed reward model, we test different combinations of the two mechanisms (i.e., the recursive mechanism and the introduction of basis functions) in the reward model. Since well-calibrated models are important for optimizing Equation~\eqref{main-problem}, we additionally introduce another metric for evaluating the reward model, which is field-level relative calibration error~\cite{Pan2019FieldawareCA}:
\begin{equation}
\text{Field-RCE}=\frac{1}{|\mathcal{D}|}\sum_{f\in\mathcal{F}}\frac{|\sum_{i\in\mathcal{D}^f}(y_i-\hat{y}_i)|}{\frac{1}{|\mathcal{D}^f|}\sum_{i\in\mathcal{D}^f}y_i},
\end{equation}
where $\mathcal{D}$ is the test dataset, $\mathcal{F}$ is the specified feature field, $\mathcal{D}^f$ is the set whose feature value is $f$, $y^i$ is the label and $\hat{y}_i$ is the predicted value. This metric measures the deviation between the predicted value of reward curve and posterior probability.

Table~\ref{table:roi_model_ablation} shows the results, and we conclude that both recursive mechanism and multi-basis functions contribute to improvement of Field-RCE and $revenue@e$.

\begin{table}[h!]
\centering
\begin{tabular}{@{}cc|cc@{}}
\toprule
\multicolumn{2}{c|}{\textbf{Variants of Reward Model}}      & \multicolumn{2}{c}{\textbf{Metrics}} 
                                        \\ \midrule
\multicolumn{1}{p{18mm}}{Recursive Mechanism} & \multicolumn{1}{p{18mm}|}{Multi-Basis Functions}  & \multicolumn{1}{c}{Field-RCE} & \multicolumn{1}{p{18mm}}{$revenue@e$} \\  
\midrule
\Checkmark                               & \Checkmark                                        & \textbf{0.137}  & \textbf{2174}  \\
\Checkmark                               & \XSolidBrush                                      & 0.148           & 2082           \\
\XSolidBrush                             & \Checkmark                                        & 0.150           & 2065           \\
\XSolidBrush                             & \XSolidBrush                                      & 0.177           & 1875           \\
\bottomrule
\end{tabular}
\caption{Results of variants of reward models under $9.96 \times 10^{12}$ FLOPs.}
\label{table:roi_model_ablation}
\end{table}

\subsection{Online Experiments}
To further demonstrate the effectiveness of GreenFlow, we conduct online A/B testing on three RSs with different levels of computational overhead, and the results are reported in Table~\ref{tab:online_result}. We observe significant decrease in FLOPs (i.e., greatly reduce the demand for model inferring servers), a slight improvement in the accuracy of RS, and negligible latency due to the introduction of GreenFlow. 
Specifically, we report the additional computation cost and latency brought by GreenFlow in Table~\ref{tab:online_result}, which can be ignored compared to the RS. Note that for RS A, the latency even decreases because the large reduction in FLOPs leads to fewer request packets, greatly reducing the time of model inference.

\begin{table}[]
\begin{tabular}{@{}c|l|ccc@{}}
\toprule
\multicolumn{2}{c|}{\textbf{Metrics}}                  & A & B & C \\ 

\midrule
\multirow{4}{*}{PFEC}            & P: Clicks             & \multicolumn{1}{r}{+2.1\%}    & \multicolumn{1}{r}{-0.2\%}     & \multicolumn{1}{r}{+0.3\%}  \\
                                 & FLOPs                & \multicolumn{1}{r}{-61\%}     & \multicolumn{1}{r}{-20\%}      & \multicolumn{1}{r}{-15\%}  \\
                                 & Energy               & \multicolumn{1}{r}{-4869kWh}  & \multicolumn{1}{r}{-168kWh}    & \multicolumn{1}{r}{-124kWh}  \\ 
                                 & $\text{CO}_2$        & \multicolumn{1}{r}{-2995kg}   & \multicolumn{1}{r}{-103kg}     & \multicolumn{1}{r}{-76kg}  \\

\midrule
\multirow{2}{*}{\shortstack{Additional\\Cost}} & Latency    & \multicolumn{1}{r}{-20ms}  & \multicolumn{1}{r}{+10ms}  & \multicolumn{1}{r}{+5ms}  \\
                                               & FLOPs      & \multicolumn{1}{r}{+3\%}   & \multicolumn{1}{r}{+8\%}    & \multicolumn{1}{r}{+8\%}   \\
                                 \bottomrule
\end{tabular}
\caption{Online A/B testing results (per day) of three RSs.}
\label{tab:online_result}
\end{table}

\begin{figure}[h!]
    \centering
    \includegraphics[width=0.41\textwidth]{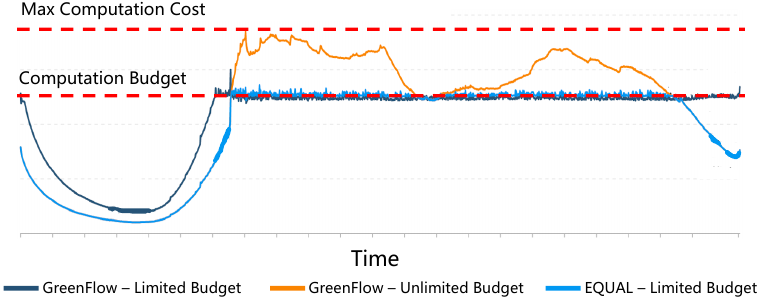}
    \caption{Computation cost of three different allocation strategies.}
    \label{fig:online_overload}
\end{figure}

Due to a large number of requests, it is also necessary to ensure that the constraint of the computation budget can be precisely satisfied in an industrial environment. When it fails, RS cannot meet all requests for recommendation and has to resort to computation downgrade, resulting in harm to revenue. In Figure~\ref{fig:online_overload}, we show computation cost of three different allocation strategies, and conclude that GreenFlow can handle traffic spikes well under the constraint of the computation budget and save a lot of computation in the same time.

To date, GreenFlow has been deployed in several RSs of an industrial mobile application with about one billion users for one year. This deployment has resulted in a significant reduction of 1,000 tons of carbon emission and a saving of approximately 2,000,000 kWh of electricity. Moreover, GreenFlow has achieved a remarkable balance among revenue, computation, carbon emission, and energy consumption.

\section{Conclusion}
In this paper, we propose a computation allocation framework for industrial RS to enhance resource utilization efficiency. The framework first estimates a reward score for each action chain, and dynamic primal-dual optimization is conducted to search for the optimal action chain while considering both the reward score and computation budget.
Through offline experiments and online A/B testing, we demonstrate the effectiveness of GreenFlow in maximizing the revenue of a RS given a particular computation budget. 
Additionally, GreenFlow has been deployed in the production environment of a mobile application, resulting in significant economic returns and a substantial reduction in carbon emissions.
In the future work, we will focus on improving the reward model since it serves as the core component that determines the computation allocation. In addition, we will make sustained efforts to open source of the core of the allocation framework with the goal of pursuing a more environmentally sound industrial RS.

\clearpage

\bibliographystyle{named}
\bibliography{ijcai23}

\end{document}